\documentstyle[aps,epsf,multicol]{revtex}
\begin{document}

\title{Dynamics of a passive sliding particle on a randomly fluctuating surface}

\author{Manoj Gopalakrishnan\thanks{e-mail: manoj@owl.phys.vt.edu}}

\address{Department of Physics and Center for Stochastic Processes in Science and Engineering,\\
Virginia Polytechnic Institute and State University,\\
Blacksburg, VA 24061-0211, USA.}

\maketitle

\begin{abstract}
We study the motion of a particle sliding under the action of an external field on a stochastically fluctuating 
one-dimensional Edwards-Wilkinson surface. Numerical simulations using the single-step model shows that the mean-square 
displacement of the sliding particle shows distinct dynamic scaling behavior, depending on whether the surface fluctuates
faster or slower than the motion of the particle. When the surface fluctuations occur on a time scale much smaller than
the particle motion, we find that the characteristic length scale shows anomalous diffusion with $\xi(t)\sim t^{2\phi}$, 
where $\phi\approx 0.67$ from numerical data. On the other hand, when the particle moves faster than the surface, its
dynamics is controlled by the surface fluctuations and $\xi(t)\sim t^{\frac{1}{2}}$. A self-consistent approximation predicts 
that the anomalous diffusion exponent is $\phi=\frac{2}{3}$, in good agreement with simulation results. We also
discuss the possibility of a slow cross-over towards asymptotic diffusive behavior. 
The probability distribution of the displacement has a Gaussian form in both the cases.

\end{abstract}

%\newpage
\begin{multicols}{2}

\section{Introduction}
The advection of a passive scalar field (such as temperature) by a turbulent fluid is a well-known 
problem\cite{PAS_SCALAR}, and is an example of a coupled semi-autonomous system, where one of the fields
evolves on it own, but affects the dynamics of the other.
Many such systems have been studied in recent times, such as the dynamics of a particle passively sliding on a randomly 
fluctuating surface\cite{DROSSEL,CHIN,DROSSEL1}, 
phase separation on a rough substrate\cite{DROSSEL1} and clustering of particles on a stochastically fluctuating surface under 
the influence of gravity\cite{DIBYENDU}. These studies have shown that such systems possess a number of non-trivial
features. For example, hard-core particles cluster into a fluctuation-dominated phase separated state\cite{DIBYENDU} 
and non-interacting particles have non-trivial density correlations\cite{DROSSEL} in the steady state. 
It is thus highly desirable 
to aim for an understanding of the relation between the time evolution of the stochastic fluctuations of the
underlying field and the dynamics of the passive scalar, and this is our primary
motivation.

In this paper, we study the dynamics of a passive sliding particle moving on a randomly fluctuating surface.
For concreteness, we choose the surface to be the 
one-dimensional Edwards-Wilkinson surface with uncorrelated noise\cite{EW}, whose 
steady state and dynamical properties are well-known. As we shall see, even this simple example has unexpectedly rich 
properties.

For clarity of presentation, we shall state the problem here. Let us consider a one-dimensional fluctuating Edwards-Wilkinson surface, 
whose equation of motion has the form

\begin{equation}
\frac{\partial h}{\partial t}=\nu \nabla^{2}h+\eta~~~~;\langle \eta(x,t)\eta(x^{\prime},t^{\prime})\rangle=
2D\delta(x-x^{\prime})\delta(t-t^{\prime})
\label{eq:EQ1}
\end{equation}

The Langevin equation for the motion of a particle {\it sliding} on the surface has the form

\begin{equation}
\frac{dX(t)}{dt}=-\Gamma \partial_{x}h|_{x=X(t)}
\label{eq:EQ3}
\end{equation}

where $X(t)$ is the position of the particle on the surface and $\Gamma$ is a friction coefficient. In principle, 
one could include a random noise term for the motion of the particle also, but since its only effect is to add a diffusive term to the 
mean square displacement, we shall drop it from further calculations. 

The time evolution of the mean-square displacement of the sliding particle is given by the equation,

\begin{equation}
\frac{d}{dt}\langle X^{2}(t)\rangle=2\Gamma^{2}\int_{0}^{t}dt^{\prime}\Phi(t,t^{\prime})
\label{eq:EQ4}
\end{equation}

where

\begin{equation}
\Phi(t,t^{\prime})=\langle \partial_{x}h[X(t),t]\partial_{x}h[X(t^{\prime}),t^{\prime}]\rangle
\label{eq:EQ5}
\end{equation}

is the 'effective' noise correlator for the motion of the sliding particle, and is equal to the slope correlator
of the fluctuating surface evaluated {\it along the particle trajectories}. The angular brackets on both side 
represents average over different realizations of the noise $\eta(x,t)$. Clearly, this is a highly
non-trivial quantity to compute, since it is not obvious {\it a priori} how the slope correlator would behave 
when averaged over such a set of highly restricted paths. 

We outline our main results here. Numerical simulations show that when the particle moves much 
faster than the surface, its dynamics is controlled by the surface fluctuations, and the mean square displacement behaves
as $\langle X^2(t)\rangle\sim \nu t$. An argument using the concept of over-turning valleys confirms this result. On the other
hand, when the surface fluctuation is fast, the particle moves in a very dynamic landscape, and the mean square displacement 
shows anomalous diffusion with $\langle X^2(t)\rangle\sim t^{2\phi}$ over the time scales of simulations. 
A self-consistent approximation
which assumes no correlation between a particle trajectory and the underlying surface height configuration, predicts
that the anomalous diffusion exponent is $\phi=\frac{2}{3}$, in reasonably good agreement with numerical results. We also
discuss the possibility that the observed anomalous diffusion might be simply transient behavior, and give reasons
for the same.

The rest of this paper is arranged in the following way. In the next section, we present the self-consistent 
approximation to evaluate the correlator in Eq.\ref{eq:EQ5}, and
discuss its predictions and limitations. In Sec. III, we present the results of numerical simulations of the problem using
the single-step algorithm. In Sec. IV, we summarize our findings and discuss the outstanding questions.

\section{The Self-Consistent Approximation}

Let us define $z=X(t)-X(t^{\prime})$, which, like $X(t)$ itself, is a random variable. Let us now sort the set of 
all available
surface configurations using $z$, i.e., for fixed $t$ and $t^{\prime}$, let $S_{z}$ denote the set of surface 
configurations
where a sliding particle is displaced by a distance $z\pm \Delta z$ between times $t$ and $t^{\prime}$. 
The fraction of such configurations is simply $P(t,t^{\prime},z) \Delta z$, where $P(t,t^{\prime},z)$ is the (unknown) 
probability distribution function for particle displacements between times $t$ and $t^{\prime}$. It then follows that

\begin{equation}
\Phi(t,t^{\prime})=\int_{-\infty}^{\infty}dz P(t,t^{\prime},z)\langle \partial_{x}h[X(t),t]\partial_{x}
h[X(t)+z,t^{\prime}]\rangle_{S_z}
\label{eq:EQ6}
\end{equation}

Note that the average in Eq.\ref{eq:EQ6} is a restricted average, i.e., it is evaluated over surface configurations 
in which the sliding particle is moved by a distance $z$ between $t$ and $t^{\prime}$. 

Up to now, our treatment has been exact. In order to make further progress, we make an approximation by assuming 
that the restricted slope correlator in Eq.\ref{eq:EQ6} is simply the standard slope correlator of the surface,
evaluated at the set of points $(0,t)$ and $(z,t^{\prime})$. In other words, we assume that the particle trajectory 
is not so strongly correlated with the surface slope configuration, such that averages of surface characteristics 
are not affected. Numerical results show that this approximation is justified except when the particle motion is 
slower than the surface fluctuations. We shall discuss this case shortly, and assume for the time being that this 
assumption is valid. This approximation is the most crucial step in our calculation. 

For the Edwards-Wilkinson surface, the slope correlator can be computed exactly, 
and the result is (for $t > t^{\prime}$),

\begin{equation}
\langle \partial_{x}h(0,t)\partial_{x}h(z,t^{\prime})\rangle=\frac{D}{\nu}\sqrt{\frac{\pi}{\nu (t-t^{\prime})}}
exp\left(-\frac{z^{2}}{\nu(t-t^{\prime})}\right)
\label{eq:EQ7}
\end{equation}

Let us now conjecture the following asymptotic scaling form for the probability distribution function of particle 
displacement:

\begin{equation}
P(t,t^{\prime},z)=\frac{1}{\xi}f\left(\frac{z}{\xi}\right)
\label{eq:EQ8}
\end{equation}

where $\xi(t-t^{\prime})$ is the characteristic dynamic length scale of particle motion. The scaling function 
$f(x)$ is required to satisfy the normalization $\int_{-\infty}^{\infty}f(x)dx=1$. It also follows that the 
mean-square displacement is 

\begin{equation}
\langle X^{2}(t)\rangle=\mu\xi^{2}
\label{eq:EQ9}
\end{equation}

where $\mu=\int_{-\infty}^{\infty}f(x)x^{2}dx$. We now substitute Eq.\ref{eq:EQ7} and Eq.\ref{eq:EQ8} into 
Eq.\ref{eq:EQ6}. The effective noise correlator thus becomes 

\begin{equation}
\Phi(t,t^{\prime})=\frac{G(T)}{\sqrt{\nu T}}~~~~;T=|t-t^{\prime}|
\label{eq:EQ10}
\end{equation}

where

\begin{equation}
G(T)=\frac{D\sqrt{\pi}}{\nu}\int_{0}^{\infty}d\eta\frac{f(\sqrt{\eta})}{\sqrt{\eta}}e^{-\frac{\xi^{2}}{\nu T}\eta}~~~; \eta=\frac{r^2}{\xi^2}
\label{eq:EQ12}
\end{equation}

We note that $G(T)$ is proportional to the Laplace transform of the function $g(x)=x^{-\frac{1}{2}}f(x)$, i.e., 
$G(T)=\frac{D\sqrt{\pi}}{\nu}{\tilde g}(\lambda$) with $\lambda=\frac{\xi^{2}}{\nu T}$.
We now conjecture a power-law growth of the characteristic length scale $\xi$ at late times:

\begin{equation}
\xi(t)\simeq  at^{\phi}~~~~~~~~~~;t\gg t_{0}
\label{eq:EQ14}
\end{equation}

where $\phi$ is the dynamical exponent that characterize the particle motion and $t_{0}$ is a microscopic time scale. 
From the preceding equations, it is clear that consistency requires $\phi >\frac{1}{2}$. For, if $\phi < \frac{1}{2}$, 
the l.h.s in Eq.\ref{eq:EQ4} would vanish at large $t$, whereas the r.h.s would not. The value $\phi=\frac{1}{2}$ is 
also inconsistent with this
equation, since the l.h.s would be a constant whereas the r.h.s would grow as $\sqrt{t}$ at large time $t$. Thus the 
only possible values are $\phi >\frac{1}{2}$. We now focus on the large-$t$ limit, where the Laplace transform 
variable $\lambda\to\infty$. The behavior of $G(T)$ in this limit follows from the following theorem\cite{SMITH}:

\begin{equation}
lim_{x\to 0}x^{-\rho}g(x)=lim_{\lambda\to\infty}\frac{\lambda^{\rho+1}{\tilde g}(\lambda)}{\rho!}~~~~;\rho > -1
\label{eq:EQ15}
\end{equation}
 
It is reasonable to assume that $f(0)$ is a non-vanishing constant, in which case $g(x)\sim x^{-\frac{1}{2}}$
as $x\to 0$. From Eq.\ref{eq:EQ15}, it then follows that $\rho=-\frac{1}{2}$, and hence
${\tilde g}(\lambda)\sim \frac{1}{\sqrt{\lambda}}$ as $\lambda\to\infty$. Thus, the asymptotic form of $G(T)$ is

\begin{equation}
G(T)\simeq \frac{\pi Df(0)}{\nu}\frac{\sqrt{\nu T}}{\xi(T)}~~~~~~; T\to\infty
\label{eq:EQ16}
\end{equation}

After substituting Eq.\ref{eq:EQ9} and Eq.\ref{eq:EQ10} into Eq.\ref{eq:EQ4}, and using the large $T$-form in 
Eq.\ref{eq:EQ16}, the resulting self-consistent integral equation for $\xi(t)$ is

\begin{equation}
\xi\frac{d\xi}{dt}=\frac{\Gamma^{2}}{2\mu}\frac{\pi D}{\nu}f(0)\int_{0}^{t}\frac{dT}{\xi(T)}
\label{eq:EQ18}
\end{equation}

We now substitute the power-law scaling form Eq.\ref{eq:EQ14} for $\xi(T)$, which gives

\begin{equation}
a^{3}\phi(1-\phi)t^{2\phi-1}\sim \frac{\pi D\Gamma^2}{2\nu}\frac{f(0)}{\mu}t^{1-\phi}~~~~;t\to\infty
\label{eq:EQ19}
\end{equation}

After equating the powers of time on both sides, we find $\phi=\frac{2}{3}$, which is the principal result of this 
paper, along with the pre-factor $a\sim\left(\frac{D\Gamma^2}{\nu}\right)^{\frac{1}{3}}$. We have thus arrived at 
the somewhat surprising result that the sliding particle moves {\it faster} than the surface fluctuations, whose 
correlation length scales with time as $\sqrt{\nu t}$. 
This super-diffusive behavior formally has its origin in the long-range nature of the 
effective noise correlator $\Phi(t,t^{\prime})$, which, from Eq.\ref{eq:EQ10} and Eq.\ref{eq:EQ16}, has a power-law 
tail of the form $|t-t^{\prime}|^{-\phi}$. Similar anomalous diffusive motion of advected particles with $\phi=\frac{2}{3}$
has been shown to occur in the one-dimensional (non-linear) Burger's equation with noise\cite{CHIN,BOHR}, using a 
mean-field approach.

Before proceeding to discuss the numerical results, we would like to point out that the calculation presented above
is valid only if $lim_{t\to \infty}\frac{\sqrt{\nu t}}{\xi(t)}$ vanishes, else we would not be able to use Eq.\ref{eq:EQ15}
to arrive at our result. Indeed, it is quite possible that in the asymptotic limit, $\xi(t)\sim \sqrt{\nu t}$, but that
this regime lies outside the validity of this approximation. In such a case, the anomalous diffusion at early times can be
still shown to exist, but with a different exponent. To see this, let us consider Eq.\ref{eq:EQ12} again at sufficiently early
times $t$ so that $\xi^{2}(t)\ll \sqrt{\nu t}$. In this case, $G(T)\simeq \frac{D}{\nu}$ is a constant, 
and after substituting in Eq.\ref{eq:EQ10},
and subsequently in Eq.\ref{eq:EQ4}, we find that $\xi(t)\sim t^{3/4}$ when $\xi^{2}\ll \sqrt{\nu t}$. From
numerical simulations (to be presented in the subsequent section), we are unable to rule out this possibility. 
Further discussions on this point are presented at the end of Sec. III.

%For ease of comparison with numerical data, it is convenient to express this result in terms of
%dimensionless variables. For this reason, we will identify the relevant units of length and time.
%For the EW surface, the parameters we have at our disposal are $D$ and $\nu$, with dimensions
%(length)$^{3}$/time and (length)$^{2}$/time respectively. From these, we identify a length scale
%${\tilde l}=\frac{D}{\nu}$ and time scale ${\tilde \tau}=\frac{D^2}{\nu^3}$. The latter is clearly
%the characteristic time scale of surface fluctuations. The particle motion is described by the parameter
%$\Gamma$, which has the dimension of velocity. Thus the characteristic time scale of particle
%motion on the surface is $\tau^{\prime}=\Gamma^{-1}{\tilde l}=\frac{D}{\Gamma\nu}$. The ratio between
%these time scales $R=\frac{\tau^{\prime}}{{\tilde \tau}}$ is a useful quantity, since this could be 
%easily changed in numerical simulations. After doing the re-scaling $\xi={\tilde l}{\tilde \xi}$, and
%$t={\tilde \tau}{\tilde t}$ where ${\tilde \xi}$ and ${\tilde t}$ are dimensionless variables, and using
%Eq.\ref{eq:EQ20}, Eq.\ref{eq:EQ14} may be written as

%$\begin{equation}
%{\tilde \xi}\sim \left(\frac{{\tilde t}}{R}\right)^{\frac{2}{3}}~~~~;{\tilde t}\gg \frac{t_{0}}{\tilde \tau}
%\label{eq:EQ21}
%\end{equation}

%Also, after equating the co-efficients on both sides, we find

%\begin{equation}
%a=\left(\frac{9}{4}\frac{\pi %D\Gamma^2}{\nu}\frac{f(0)}{\mu}\right)^{\frac{1}{3}}
%\label{eq:EQ21}
%\end{equation}

\section{Numerical Results}

In this section, we discuss the results of numerical simulations of the problem. The surface is constructed as a set of 
height 
variables $\{h_{i}\}$, where $i=1,2,..N$ and $N$ is the lattice size. Periodic boundary conditions are imposed on 
the surface. We use the single-step model of the surface\cite{SOS}, where the height difference between adjacent 
lattice points is restricted, i.e., 
$\tau^{\pm}_{i}=\pm[h_{i\pm1}-h_{i}]=\pm 1$, 
where we have defined the clockwise/anti-clockwise slope variables $\tau^{\pm}_{i}$. 
The surface evolves through the 
following set of dynamical rules: 

\begin{itemize}
\item If $\tau^{+}_{i}=1$ and $\tau^{-}_{i}=-1$, $h_{i}\to h_{i}+2$ with probability $q$ ($q\leq 1$).\\

\item If $\tau^{+}_{i}=-1$ and $\tau^{-}_{i}=1$, $h_{i}\to h_{i}-2$ with probability $q$.

\end{itemize}

The large-distance, long-time properties of this model are identical to the continuum model in Eq.\ref{eq:EQ1}.
We start from a 'ordered' height configuration, where $h_{i}=1$ is $i$ is odd, and $h_{i}=0$ when $i$ is even. A 
lattice site is selected at random, and its height is updated in accordance with the above rules. 
For the surface configuration, one Monte Carlo step is counted after $N$ such 
attempted updates, where $N$ is the lattice size. The surface then evolves through $T\simeq 10N^2$ Monte Carlo time 
steps in order to reach the steady state. We used two lattice sizes, $N=1024$ and $N=4096$ in our simulations. 

For a certain surface configuration in the steady state, we place 100 particles on the surface at randomly selected 
locations. At 
each Monte Carlo step, after a surface update is complete, a particle is selected at random (with initial position 
at lattice 
site $i$, say) and its position is updated according to the following set of dynamical rules.

\begin{itemize}
\item If $\tau^{+}_{i}=1$ and $\tau^{-}_{i}=1$, the particle moves one lattice step in the clockwise 
direction with probability $p$ ($p\leq 1$).\\
\item If $\tau^{+}_{i}=-1$ and $\tau^{-}_{i}=-1$, the particle moves one lattice step in the 
anti-clockwise direction with probability $p$.\\
\end{itemize}

Essentially, the rules say that the particle only slides down when the slope is favorable, and remains stationary
when it is at a local minimum or maximum. The particle displacements are measured such that for every move in clockwise
direction, $X(t)\to X(t)+1$, and for every move in anti-clockwise direction, $X(t)\to X(t)-1$.
The particles are assumed non-interacting, which is consistent with the 
single-particle picture. 
When the probability $p > q$, the particles moves fast relative to the surface motion and {\it vice-versa}. Also, without 
any loss of information, we set max$(p,q)=1$. The parameter we tune in the simulations is the ratio $r=\frac{p}{q}$. 
The simulations were run up to $10^{5}$ Monte Carlo steps for $N=1024$ and $10^6$ Monte Carlo steps for $N=4096$.
We computed the probability distribution of the displacement of the particles, 
as well as the mean square displacement, as  functions of time. The results were averaged over 100 different 
starting configurations for $N=4096$ and 1000 starting configurations for $N=1024$.

In Fig.1, we have shown the results for the RMS displacement of the sliding particle plotted against the quantity 
$qt$, where $t$ is the number of Monte Carlo
steps. The fraction $q$ is used to scale the time so that all the surface configurations pass through an almost equal
number of updates (In the continuum language, this is equivalent to using the combination $\nu t$). We have shown the results for
four values of the update frequency ratio $r$: 0.01, 0.1, 1 and 10.
The results for $r=0.01$ and $r=0.1$ offers evidence for the power-law behavior $\xi\sim t^{\frac{2}{3}}$ 
as predicted by the self-consistent argument. For $r > 1$, on the other hand, the growth of the RMS displacement with time 
is much slower. In this case, the observed growth exponent is closer to $\frac{1}{2}$ at all times. The case $r=1$ displays
intermediate behavior, with an effective  exponent close to 0.56.

\begin{figure}
\epsfxsize=1.8in
\epsfbox{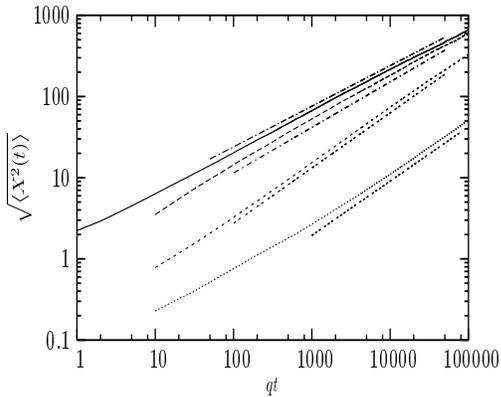}
\narrowtext
%\vspace{-2.0cm}
\caption{The figure shows the time evolution of the root-mean-square displacement of the sliding particle for several
values of the update frequency: $r=10, 1,0.1$ and 0.01 (top to bottom). The short straight lines are guides to the
eye, and have slopes 0.5, 0.56, 0.68 and 0.67 (top to bottom). 
When $r<1$, the dynamic exponent is close to 2/3, whereas when $r > 1$, it is close
to 1/2. For $r=0.01$, there is an early-time regime where the growth exponent is smaller than 2/3, but the
asymptotic behavior is identical to that of $r=0.1$. The lattice size in these simulations is $N=4096$.}
\end{figure}

Although it is tempting to say that the cases $r>1$ and $r<1$ are characterized by different dynamical exponents, we would
like to be cautious here. Our numerical data does not rule out the possibility that the observed anomalous diffusion
of the sliding particle when $r< 1$ is simply a transient regime, and the asymptotic behavior might be identical for
all values of $r$. In the scenario discussed in the last paragraph of the previous section, the observed behavior might
reflect a slow cross-over from $\xi\sim t^{3/4}$ to $\xi\sim t^{\frac{1}{2}}$.
The data for $r=0.1$ shows some bending for $t > 10^5$, but at the moment we cannot conclusively
say whether this shows the cross-over towards a diffusive regime, or simply a finite size effect. We have checked that
a smaller system size ($N=1024$) shows bending as early as $t\simeq 10^4$ (Fig.2), which makes it likely that
this is a finite size effect.

\begin{figure}
\epsfxsize=1.8in
\epsfbox{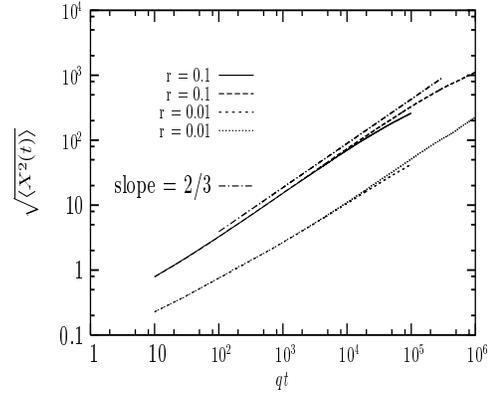}
\narrowtext
%\vspace{-2.0cm}
\caption{The figure illustrates finite size effects in the time evolution of the root-mean-square displacement of the 
sliding particle for $r=0.1$ (top) and $r=0.01$ (bottom), for lattice sizes $N=1024$ and $N=4096$. The short straight 
line at the top is a fit, and has slope 0.67.
}
\end{figure}

{\it Particle trapping in the valleys:} The origin of the distinct scaling behavior when $r>1$ 
may be understood through the following argument. When the particle moves faster than the underlying surface 
fluctuations, 
after the initial downward slide towards the local minimum, it gets 'trapped' there. At all future times, the 
motion of the 
particle is 'slaved' to the dynamics of this local minimum. 
The long-time dynamics of the particle may be visualized using the concept of {\it valleys} in the 
surface. 
A valley is defined as a linear stretch of the surface below a certain reference height, which we choose to be the 
instantaneous height at 
the starting location of the particle (Fig. 3). It is well-known that any instantaneous height configuration of a EW 
surface relative to a 
reference point on the surface is a random walk (RW), which returns to the origin after $N$ steps. Thus, the length 
distribution of 
the valleys is simply the distribution of return times of a one-dimensional RW to its starting point, which follows a 
power-law decay: 
$P(l)\sim l^{-\frac{3}{2}}$ for $l\ll N$\cite{FELLER}. 
Let us now consider a section of the surface with length $\xi$, and let $N_{\xi}(l)$ be the number of valleys of 
length $l$ 
within this section. Since we require $\int dlN_{\xi}(l)l=\xi$, it follows that $N_{\xi}(l)$ has the scaling form

\begin{equation}
N_{\xi}(l)\sim \sqrt{\xi}l^{-\frac{3}{2}}
\label{eq:EQ20}
\end{equation}  

The slope correlations of the surface over a length $l$ will decay by a time $\tau(l)\sim l^{2}/\nu$ from
Eq.\ref{eq:EQ7}. For a valley of length $l$, this time is roughly the time scale for the {\it over-turning} 
of the valley, whereby the valley is transformed into a hill. An event of over-turning leads to a displacement of a particle 
trapped inside the valley by a distance $\sim l$. The total time required to produce a displacement $\sim \xi$ is then the sum of 
the trapping times inside all the valleys within $\xi$, which is given by

\begin{equation}
t(\xi)\simeq \int_{0}^{\xi}dl\tau(l)N_{\xi}(l)
\label{eq:EQ21}
\end{equation}
 
After using Eq.\ref{eq:EQ20}, this equation gives $t(\xi)\sim \xi^{2}/\nu$. Equivalently, for a fixed time $t$, the typical
displacement of the sliding particle is $\xi(t)\sim \sqrt{\nu t}$. The result shows that in the regime $r > 1$, 
the dynamical exponent for the sliding particle is $\frac{1}{2}$.

\begin{figure}
\epsfxsize=1.8in
\epsfbox{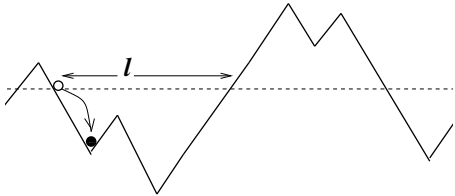}
\narrowtext
\vspace{-2.0cm}
\caption{ Illustration of a valley in the single step model: The open circle shows the position of
the particle at time $t=0$ and the filled circle shows its current position. The horizontal 
straight line divides the surface into valleys and hills.}
\end{figure}

\begin{figure}
\epsfxsize=2.1in
\epsfbox{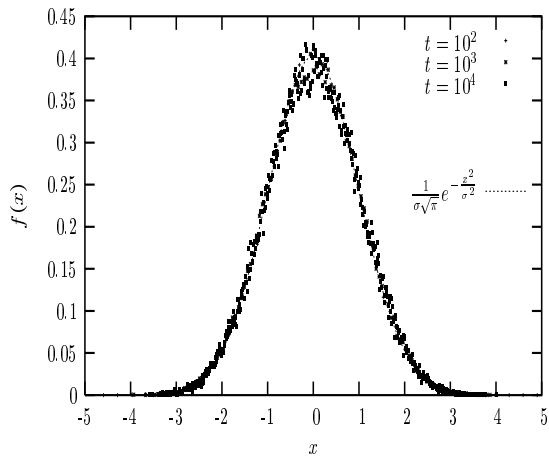}
\narrowtext
\vspace{-1.0cm}
\caption{ 
The figure shows the scaled probability distribution $f(x)={\langle X^{2}(t)\rangle}^{\frac{1}{2}}P(0,t,z)$ plotted 
against the 
scaled distance $x=r/\sqrt{\langle X^{2}(t)\rangle}$ for three widely separated values of time. The good scaling 
collapse justifies the scaling form used in Eq.\ref{eq:EQ8}. The lattice size is $N=1024$ and the update
frequency ratio is $r=0.1$ in this figure. The parameter $\sigma\approx 1.45$ in the Gaussian fit, and is found
to be the same for all values of $r$. The data represents an average over 1000 different surface starting configurations.
}
\end{figure}

Finally, we discuss the results for the probability distribution $P(0,t,z)$ of the displacement of the particle at time
$t$, relative to its location at time $t=0$. In Fig.4, we have displayed the results for $r=0.1$ at three widely separated 
instants of time, $t=10^2, 10^3$ and $10^4$. When the scaling function $f(x)={\langle X^{2}(t)\rangle}^{\frac{1}{2}}P(0,t,z)$ of the 
probability
distribution is plotted against the scaled distance $x=z/\sqrt{\langle X^{2}(t)\rangle}$, we find good collapse 
of the data, which provides a {\it a posteriori} justification for using this scaling form. It is also remarkable
that the scaling function is very well represented by a Gaussian (see the fit in the figure).

\section{Conclusions}
To conclude, we have shown by means of a self-consistent argument that the motion of a passive sliding particle on a 
fluctuating
surface may show non-trivial behavior even when the surface is Gaussian, like the Edwards-Wilkinson surface. 
Specifically, the effective dynamical exponent for the motion of the sliding particle may be the same as, or 
different from the dynamical
exponent of the surface fluctuations, depending on the relative 
time scale of surface and particle motion. The observed apparent super-diffusive dynamics of particle displacement in the 
EW surface 
urges us to be cautious when the measurements of the passive sliding particle is used to
measure the dynamical exponent or other dynamical properties of the surface itself, as has been suggested in \cite{CHIN}. 
It might be interesting to extend this study to look into the effects of changing the relative update frequency of the 
particle and surface configurations on the steady state 
characteristics of the problem in a many-particle context\cite{DROSSEL,DROSSEL1}. Lastly, simulations on a larger time
scale using much bigger lattices would be necessary to determine conclusively if the observed anomalous diffusion is
indeed a cross-over effect.

\section{Acknowledgments}

This work has been supported in part by a grant (DMR-0088451) from the U.S. National Science Foundation. 
The author would like to thank B. Schmittmann for valuable discussions and 
B. Drossel and R. K. P. Zia for critical comments and suggestions.

\end{multicols}

\end{document}